\documentclass{natureH}

\usepackage{amsmath}
\usepackage{amssymb}
\usepackage{graphicx}
\usepackage{accents}
\usepackage{etoolbox}
\usepackage{hyperref}
\usepackage{pdfpages}
\usepackage{setspace}
\usepackage{caption}

\makeatletter
\let\saved@includegraphics\includegraphics
\AtBeginDocument{\let\includegraphics\saved@includegraphics}
\renewenvironment*{figure}{\@float{figure}}{\end@float}
\makeatother

\newcommand{\Brk}[1]{\left[ #1 \right]}
\newcommand{\BRK}[1]{\left\{ #1 \right\}}

\newcommand{\abs}[1]{\left| #1 \right|}

\newcommand{\dg}{^\circ}
\newcommand{\abar}{\bar{a}}

\newcommand{\sfig}{m} 
\newcommand{\SI}{Supplementary materials}

\title{Making Faces: Universal Inverse Design of Surfaces with Thin Nematic Elastomer Sheets}
\author
{Hillel Aharoni,$^{1\dagger}$ Yu Xia,$^{2\dagger}$, Xinyue Zhang$^{2}$, Randall D. Kamien$^{1}$, Shu Yang$^{2\star}$}
\date{}

\begin{document}

\baselineskip24pt
\maketitle
\begin{affiliations}\footnotesize
\item Department of Physics and Astronomy, University of Pennsylvania, Philadelphia, PA 19104, USA
\item Department of Materials Science and Engineering, University of Pennsylvania, Philadelphia, PA 19104, USA
\item[] $^\dagger$These authors contributed equally to this work.
\item[] $^\star$E-mail: shuyang@seas.upenn.edu.
\end{affiliations}

\begin{abstract}
	Programmable shape-shifting materials can take different physical forms to achieve multifunctionality in a dynamic and controllable manner. Although morphing a shape from 2D to 3D via programmed inhomogeneous local deformations has been demonstrated in various ways, the inverse problem -- programming a sheet to take an arbitrary desired 3D shape -- is much harder yet critical to realize specific functions. Here, we address this inverse problem in thin liquid crystal elastomer (LCE) sheets, where the shape is preprogrammed by precise and local control of the molecular orientation of the liquid crystal monomers. We show how blueprints for arbitrary surface geometries as well as local extrinsic curvatures can be generated using approximate numerical methods. Backed by faithfully alignable and rapidly lockable LCE chemistry, we precisely embed our designs in LCE sheets using advanced top-down microfabrication techniques. We thus successfully produce flat sheets that, upon thermal activation, take an arbitrary desired shape, such as a face. The general design principles presented here for creating an arbitrary 3D shape will allow for exploration of unmet needs in flexible electronics, metamaterials, aerospace and medical devices, and more.
\end{abstract}
\pagebreak


A common modality of biological systems is the ability to shape themselves into various morphologies in order to interact with their surrounding environment, a concept also widely adopted by mankind in various technological contexts\cite{Burgert2009}. Much effort has been invested in the last decades in the research of soft systems, that deform smoothly and continuously, often inspired by natural systems. Examples of such studies include, to name just a few, environmentally responsive hydrogels \cite{KES07,Kim2012}, self-folding \textit{origami} \cite{Tolley2014, Na2015} and \textit{kirigami} \cite{Zhang2015}, or smart textiles \cite{Hu2012}, with applications in soft robotics \cite{Shepherd2011,Trivedi2008}, biomedicine \cite{Woltman2007,Fernandes2012,Randall2012}, functional and artistic design \cite{Ritter2007}. In natural systems, the local structure often appears to be highly optimized for obtaining a certain shape that is necessary for fulfilling a certain function. Such high-end design is hard to achieve artificially, and biomimetics is often used to reproduce designs that appear in the natural world \cite{Ionov2013,SydneyGladman2016}. Still the question remains, how does one create a new stimulus-responsive design from scratch? How does the local structure at every point need be set in order to get the desired global shape and response?

The answer depends on the local properties of the materials in use. For example, thin hydrogel sheets with programmable local isotropic expansion are widely studied in the context of shaping \cite{KES07,Kim2012}. In these systems, the inverse design problem is mathematically well understood and generally solved \cite{Chern1955}. However, properties such as low elastic modulus, small deformations, slow response times, or the need for very specific aqueous environmental conditions may render such systems unsuitable for many applications. A class of materials that exhibit properties that are well suited for a wide range of design applications is liquid crystal elastomers (LCEs), that are lightly crosslinked polymer networks of liquid crystals (LCs). The inherent molecular anisotropy of LCs allows engineering the mechanical response of LCEs with fine control over both the magnitude and direction of the local strain. Compared to osmosis-driven hydrogel-based responsive materials, LCEs are characterized with high mechanical strength and quick response ($\sim 100\,ms$) \cite{Camacho-Lopez2004}, while exhibiting large anisotropic local deformations ($\sim 300\%$) in response to a variety of external stimuli, e.g. heat, light, or magnetic field.
These materials have therefore been in the focus of much research in recent years, both theoretical and experimental \cite{Warner2007}. Since the anisotropic local deformation is different from the isotropic one, describing the local geometry of LCEs requires an entirely different mathematical framework than that used to describe isotropic materials systems such as hydrogels. Namely, it requires linking the geometry to the local molecular orientation -- the nematic director -- which is a principal axis of the local anisotropic deformation. Several recent theoretical works establish the relation between the nematic director field in an undeformed LCE sheet and the geometry it adopts upon stimulation, either in a discrete framework \cite{Modes2010,Modes2011} or in a continuous one \cite{aharoni2014geometry, Mostajeran2016}. 

The inverse design problem for LCE-type deformations, i.e. finding a planar director field that would induce an arbitrary 2-dimensional (2D) geometry, was discussed in \cite{aharoni2014geometry} and later in \cite{SydneyGladman2016} but was not solved for the general case. General solutions were found for similar, less constrained inverse problems, namely when relaxing the planar-field demand and assuming full control over the 3-dimensional (3D) director field at every point \cite{Plucinsky2016}, or when relaxing the fixed magnitude demand and assuming full control over the entire planar deformation tensor \cite{VanRees2017}. 
Nonetheless, experimental realization of those assumptions will be extremely challenging. This is in contrast to planar-director LCEs, for which many experimental methods have been investigated, aiming to improve the accuracy and spatial resolution of the nematic director field at every point. Initial attempts to construct both smooth curvature fields and discrete ones proved successful \cite{DeHaan2012,Mostajeran2016,Ware2015,Xia2016},
thus laying the cornerstone for realizing arbitrary designs of thin LCE sheets. Nevertheless, there remain several obstacles before reaching this goal; first, the general inverse design problem for planar fields is still unresolved; second, assuming a solution exists, one needs to be able to accurately and faithfully imprint it onto a thin LCE sheet and obtain the correct local stimulus-responsive deformation in practice; third, obtaining the right 2D geometry is not enough since it can have many different isometric embeddings in space, and since the system might end up being dynamically stuck in a non-isometry or in an otherwise undesired configuration. Therefore, control over additional degrees of freedom needs be achieved in order for these methods to be applied for general designs.

\begin{figure}
	\centering
	\includegraphics[width=.8\columnwidth]{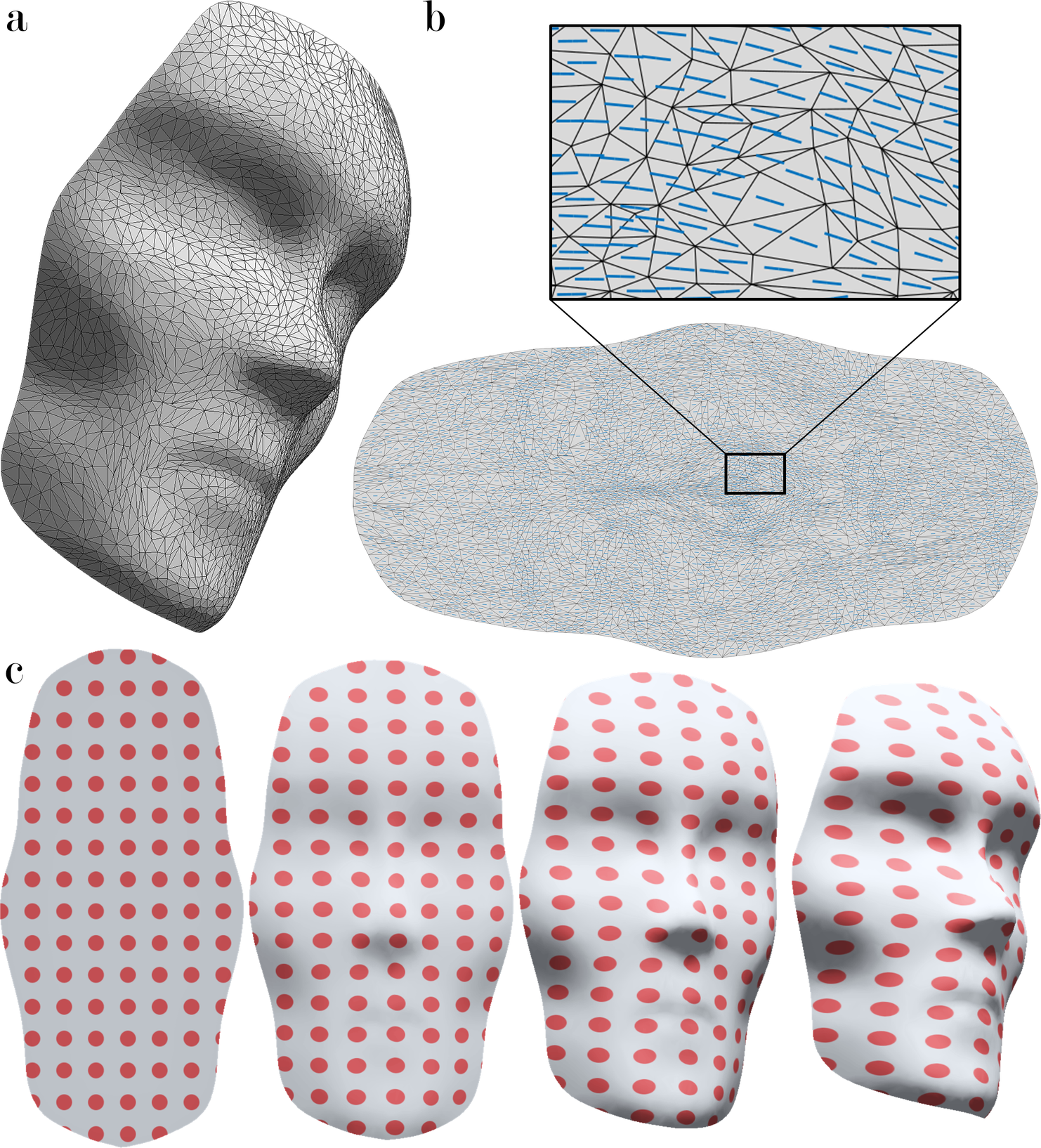}
	\caption{Numerical inverse design of arbitrary geometry.
		(a) The input for our numerical algorithm is a surface in 3D, given in the form of a triangulated mesh.
		(b) As output, our algorithm assigns to each triangle 2D coordinates and a director angle (blue segments). The deformation between every plane-triangle and its matching surface-triangle approximates the anisotropic material deformation at the target temperature. This deformation reflects, at the target temperature, the 2D geometry given by eq.~(\ref{metric}).
		(c) The deformation visualized with Tissot's indicatrices. Circles in the undeformed planar domain (left) map onto the deformed surface as ellipses, with constant temperature-dependent principal axes $\BRK{\lambda, \lambda^{-\nu}}$ and with principal directions aligned along the calculated local director. At the target $\lambda$, the desired surface is reconstructed (right).}
		\label{fig1}
\end{figure}

In this work we resolve some of these issues in a first attempt to allow the design of arbitrarily shaped surfaces from flat LCE sheets, implementing several novel theoretical and experimental techniques. We first take a numerical approach in approximating a solution to the inverse design problem, thus obtaining a planar nematic director field that equips the sheet with the desired 2D metric tensor. We realize these designs using a variation on the top-down microfabrication technique that was previously developed by some of the authors \cite{Xia2016} to precisely control the molecular orientation at every point in the LCE sheet. Combined with new LCE polymer chemistry based on a ``click'' reaction that can efficiently and faithfully produce uniform strain across the film, we create a high-accuracy realization of the theoretical designs with precise control over both the direction and magnitude of the local strain in the polymer films. Furthermore, we incorporate higher-order techniques, namely the introduction of local target directional curvatures throughout the sheet and specifically around its edge, in order to better insure buckling of the sheet towards the correct isometry and avoid stagnation at local equilibria away from the desired design. We show how the combination of these methods allows us to create sheets that upon activation take an arbitrary design, e.g. a face (Fig.~\ref{fig1}).

The mathematical formulation of the inverse design problem for thin planar nematic elastomer plates was presented in \cite{aharoni2014geometry}. It involves, given a desired 2D metric tensor $\abar$, finding coordinates $\BRK{u,v}$ and a director field $\theta(u,v)$, so that $\abar$ takes the form
\begin{equation}\label{metric}
\abar(u,v)=R\Brk{\theta(u,v)}\begin{pmatrix}
\lambda^2&0\\0&\lambda^{-2\nu}\end{pmatrix}R\Brk{\theta(u,v)}^T.
\end{equation}
Here, $\lambda$ and $\lambda^{-\nu}$ are the local expansion ratios parallel and perpendicular to the nematic director field respectively, and $R[\theta]$ is the planar rotation matrix by angle $\theta$.

\begin{figure}
	\centering
	\includegraphics[width=.8\columnwidth]{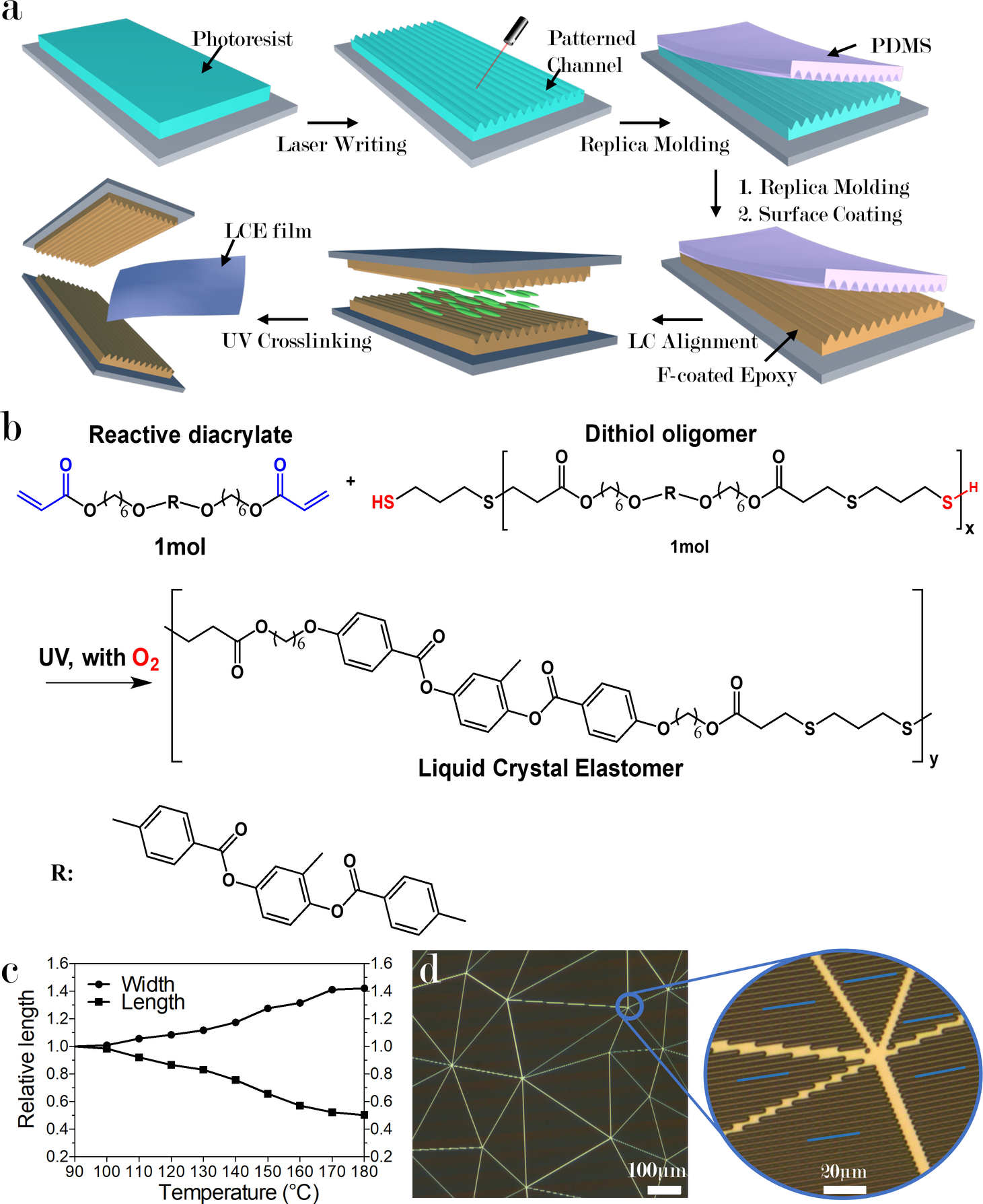}
	\caption{Achieving accurate control over local deformation.
		(a) Experimental procedure.
		(b) Schematic of the thiol-acrylate ``click'' reaction for precise control of local strain in LCEs.
		(c) Deformation of uniformly aligned free LCE film, in the directions parallel and perpendicular to the director, as a function of temperature.
		(d) An optical microscope image of channels patterned onto the epoxy coated glass mold for controlling the local alignment of the planar nematic director field. Individual channels are visible in the further magnified inset, where channel orientation in each monolithic domain is denoted by the blue segments.}
	\label{fig2}
\end{figure}

We take a numerical approach for finding an approximate solution to eq.~(\ref{metric}). We start with a desired surface given as a triangulated mesh in 3D, whose vertices are initially parameterized by some 2D coordinates, and with certain values of $\lambda$ and $\nu$ that describe the local material deformation at a certain desired temperature (Fig.~\ref{fig2}c). Following the approach of \cite{aharoni2014geometry}, we iteratively modify this coordinate parameterization to bring the transformed metric tensor's eigenvalues close to $\BRK{\lambda^2,\lambda^{-2\nu}}$ at every mesh triangle. Our algorithm for this procedure, based on a numerical algorithm presented in \cite{Liu2008}, is described in detail in \SI. Its output is a 2D coordinate mesh, $\BRK{u,v}$, with the same triangulation as our input surface, along with a director angle $\theta$ defined at each mesh triangle. The 2D intrinsic geometry of such a sheet is expected to follow Eq~(\ref{metric}), hence it is Euclidean at preparation temperature ($\lambda_{\rm{prep}}=1$) and approximates the geometry of our desired surface upon accepting the target values of $\lambda$ and $\nu$. We evaluate the approximation error, and refine the mesh or run the algorithm starting at a different initial condition if necessary.


To realize the designs given by our numerical algorithm, it is crucial to prepare spatially heterogeneous LCEs with precisely controlled lateral deformation at every point. It thus requires control not only of the arbitrary position-dependent orientation of the nematic director, but also of the magnitude of local principal strains throughout the film. Since the theory and numerical algorithm assume constant principal strains $\BRK{\lambda^2,\lambda^{-2\nu}}$ across the film, it is necessary to fulfill this requirement in the real experimental system. However, traditional LCE chemistry\cite{Broer2011} that is based on free radical polymerization seriously suffers from the inhibition reaction of oxygen from the ambient air, and the uniformity of crosslinking density in the resultant LCE film cannot be guaranteed, which in turn would affect the local strain in the film. To overcome this challenge, we develop a novel surface alignable LCE chemistry based on an oxygen mediated thiol-acrylate ``click'' reaction (Fig.~\ref{fig2}b), allowing for immediate locking of the LC director field in the local domains. In thiol–acrylate reactions, it has been reported\cite{Cramer2001,OBrien2006} that oxygen serves as a chain transfer agent that promotes the condensation reaction of thiol-acrylate, and the polymerization rate is not affected significantly by the presence of oxygen (see reaction mechanism in Fig. S1). Seeking to enable high and uniform strain in LCE, we take advantage of this oxygen-friendly ``click'' reaction, and use bifunctional dithiol and diacrylate as precursors to produce main-chain LCEs (Fig.~\ref{fig2}b).

Unfortunately, mixing high fraction of non-mesogenic reactant into mesogenic liquid crystal could create a new problem in maintaining the LC phase \cite{Ware2015a}, as a result of the drastic influence of the non-mesogenic chains on the LC packing symmetry. To suppress this effect, we prepare LCEs via a two-step synthetic procedure. First, we pre-polymerize mesogenic diacrylates (RM82) and non-mesogenic dithiol (1,3-propane dithiol) via a base-catalyzed ``click'' reaction to generate a new mesogenic dithiol oligomer (Fig. S2).
After reaction, the dithiol oligomer is in a nematic phase at room temperature and can be homogenously mixed with RM82 at any fraction ratio without destroying the nematic phase. The mixture has an extremely large nematic window of $\sim118\dg C$ (Fig. S3a). Finally, we obtain the LCEs through photopolymerization of the RM82/dithiol oligomer mixture via "click" chemistry (Fig. S1a), which is found extremely efficient, leading to $85\%$ gelation after 20 seconds of UV exposure at $50~mW/cm^2$ (Fig. S4a).

After crosslinking, we end up with a low modulus and low glass transition temperature elastomer (Fig. S3b,S4b). The LCE film shows clear birefringence and shape changes in response to heating (Fig. S5a). We prepare a uniformly aligned LCE film to measure the variation in the length and width of the film upon heating, and calculate the thermo-responsive strain plotted in Fig.~\ref{fig2}c and Fig. S5b. The maximum strain of our LCE system is found to be $\sim 100\%$ in the system of RM82/dithiol oligomer at $1:1$ ratio of functional group. These large values of $\lambda$ grant access to a wide variety of 2D geometries via our numerical algorithm.

\begin{figure}
	\centering
	\includegraphics[width=.8\columnwidth]{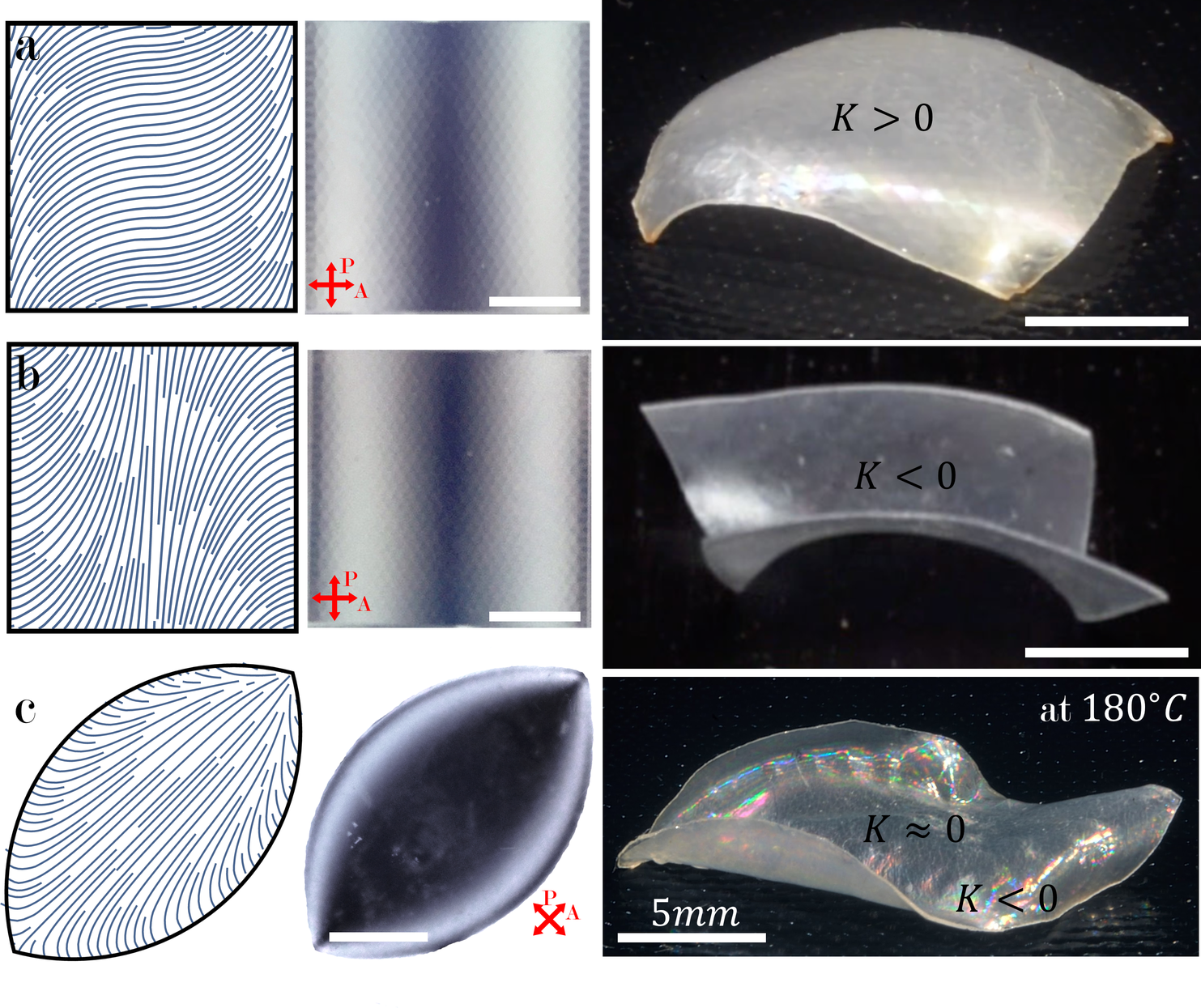}
	\caption{Designing simple shapes using only their 2D geometry.
		The theoretically or numerically obtained planar director fields (left) are imprinted onto the LCE sheet.
		We measure the experimentally obtained director field using crossed polarizers to verify that it coincides with the desired field (middle).
		When brought to the target temperature $180\dg C$ (right), the initially flat LCE sheet takes the desired 3D shape.
		The procedure is shown for the following geometries:
		(a) Constant positive Gaussian curvature,
		(b) Constant negative Gaussian curvature, and
		(c) Leaf-shaped surface, with strong negative Gaussian curvature around the sides.
		All scale bars are $5mm$.}
	\label{fig3}
\end{figure}

Spatial control over the nematic director field has been demonstrated in photo-alignment systems\cite{Ware2015} and in our previously reported photo-patterned 1D channel system \cite{Xia2016} (with feature size $2\mu m$). Here, we implement our method to align LCs according to our numerical designs. To further improve the alignment quality, we employ a new photo-patterning technique through direct laser writing (DLW, see details in \SI) to prepare even smaller pattern feature size of $1\mu m$ that gives a much higher surface anchoring energy \cite{Berreman1972} (detailed discussion in \SI). We prepare channel patterned molds with channel width of $1\mu m$, spacing of $1\mu m$, and depth of $600 nm$, arranged into monolithic domains of average size $\sim 200-500\mu m$ (Fig.~\ref{fig2}d, S6), which correspond with the discrete elements in our numerical algorithm. The two sides of the a mold are placed with a $150 \mu m$ spacing between them, into which we insert the RM82/dithiol oligomer melt. The nematic director locally aligns with channel direction, and fast UV crosslinking traps it inside the solid LCE films (Fig.~\ref{fig2}a, S4a). We use this system to realize designs obtained by our numerical algorithm, with good success in reproducing several simple designs. Some of these are shown in Fig.~\ref{fig3} -- surfaces of constant positive and negative Gaussian curvatures (which can be thought of as building blocks for arbitrary 2D geometries), and the shape of a leaf with negative Gaussian curvature exponentially decaying away from the edge \cite{SRS07}.

Nonetheless, we face more challenges when trying to produce more complex models, e.g. the face in Fig.~\ref{fig1}, whose 2D geometry consists of smaller length scales and alternating curvatures. One reason for this might be that with such geometries the thin limit assumption is less valid and bending energy plays a bigger role trying to locally flatten the sheet.
Considering the size of our monolithic domains $\sim 200-500\mu m$ as a lower bound on the radii of curvature in our designs, we alleviate this problem by decreasing the film thickness to $100\mu m$ to approach the thin limit, however with the mechanical integrity of the LCE film in mind, we refrain from going lower. Another problem with complex designs is the abundance of different isometries and near-isometries of the desired shape, which is further aggravated by the dynamic process in which different regions in our sample might initially buckle in different directions and be left stuck away from the system's global ground state.

To grapple with these issues we augment our designs with a small gradient in the nematic director across the sheet's thickness, introducing non-trivial local reference curvatures. As is explained in \cite{aharoni2014geometry}, this method cannot be used to create any desired local curvature; with strictly planar anchoring, it can only induce a reference curvature tensor with equal and opposite principal curvatures, with principal curvature directions at $\pm 45\dg$ with respect to the thickness-averaged director. The only additional scalar-field degree of freedom is the absolute value of these principal curvatures, which is proportional to the magnitude of the local director field gradient at every point \cite{aharoni2014geometry}. We calculate the value of this scalar field $\Delta\theta(u,v)$ that minimizes the discrepancy between the induced reference curvature and the curvature of the desired field. A detailed calculation can be found in the \SI. The output of the numerical calculation are values of $\Delta\theta$ evaluated at each mesh triangle. This gives us for each mesh triangle two director fields: $\theta+\Delta\theta/2$ for the top surface of the sheet and $\theta-\Delta\theta/2$ for the bottom surface. Since the difference between these two fields is small (in our models $\abs{\Delta\theta}\lesssim 5\dg$) and continuous, the nematic strain caused by it is not enough to compromise surface anchoring or to promote nematic defects, nor does it significantly change the induced reference metric (see discussion in \SI).

\begin{figure}
	\centering
	\includegraphics[width=.8\columnwidth]{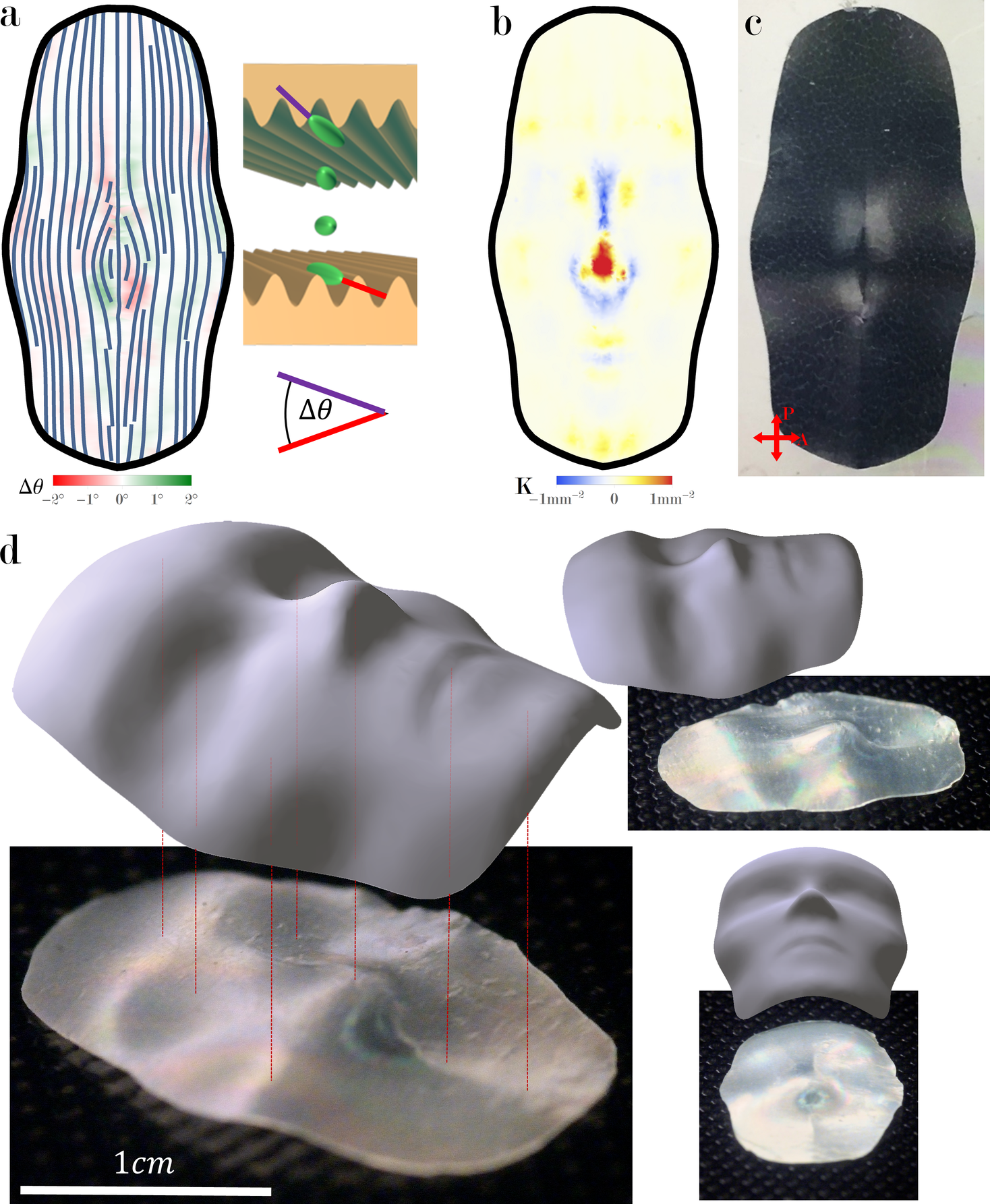}
	\caption{Designing a face using both metric tensor and curvature tensor.
			(a) Two slightly different director fields are imposed on the top and bottom surfaces of the LCE sheet. Shown streamlines indicate the director field at the mid-plane, while color indicates the angle difference $\Delta\theta$ between the top and bottom (illustrated by inset sketch).
			(b) Gaussian curvature of the target surface. The correlation between regions with strong negative Gaussian curvature and regions with strong $\abs{\Delta\theta}$ stems from the saddle-like form of the induced curvature tensor. 
			(c) Polarized optical microscopy is used to verify the resulting director field throughout the LCE sheet.
			(d) At the target temperature $180\dg C$ the LCE sheet takes the shape of a face.}
	\label{fig4}
\end{figure}

We realize such a director field by creating two independent channel patterns for the top and bottom parts of the mold, instead of using two mirror images of the same pattern. We still use the same triangulated mesh for both sides, only the channels within each monolithic domain are oriented in slightly different directions (Fig.~\ref{fig4}a). The resulting director field remains essentially planar throughout the sample, and twists smoothly from top to bottom to match the direction of the channels on both sides. When heated, the reference curvatures cause the resulting LCE film to buckle in the desired direction at certain areas, thus pushing it dynamically toward the preferred isometry -- a face (Fig.~\ref{fig4}d and Fig. S7).


This work brings together novel numerical and experimental methods to address a universal inverse problem, that is, how to reconstruct 3D surfaces from 2D sheets of anisotropic LCEs. We established an explicit protocol for preprogramming any desired shape into an LCE sheet.
Our protocol is not limited in material or scale; liquid crystals are not necessarily molecular, they can be any material with long range orientational order, and an ``LCE-like'' anisotropic material could be any liquid crystal coupled with any elastic medium. Our designs can therefore be broadly used, in systems ranging from cellulose nanocrystals embedded in polymer networks to sewing threads embedded in hydrogels. This allows the design of shape-shifting surfaces at a wide variety of length scales and material properties, which, combined with the shape-universality of our design scheme, opens the door for countless technological applications.


\begin{addendum}
	\item We acknowledge support by the National Science Foundation (NSF) DMR/Polymer program, DMR-1410253 (S.Y.). This work is also partially supported by NSF/EFRI-ODISSEI EFRI 13-31583, NSF DMR-1262047, and a Simons Investigator grant from the Simons Foundation (R.D.K.).
	\item[Additional Information] Supplementary information is available in the online version of the paper.
\end{addendum}




\section*{Supplementary material}

# Supplementary Materials for

## Making Faces: Universal Inverse Design of Surfaces with Thin Nematic Elastomer Sheets

Hillel Aharoni, Yu Xia, Xinyue Zhang, Randall D. Kamien, and Shu Yang

**This document includes:**

Materials and Methods

Supplementary Text

References S1 to S7

Figures S1 to S7

Captions for Movies S1 to S5

1

**Materials and Methods**

Numerical Inverse Design of 2D Metric Tensor

The inverse design problem for thin planar nematic elastomer sheets was presented by Aharoni *et al.* [S1]. It involves, given a desired two-dimensional metric tensor $\bar{a}$, finding coordinates $\{u, v\}$ in which $\bar{a}$ takes the form

$$\bar{a}(u,v) = R[\theta(u,v)] \begin{pmatrix} \lambda^2 & 0 \\ 0 & \lambda^{-2\nu} \end{pmatrix} R[\theta(u,v)]^T \qquad (S1)$$

for some director field $\theta(u,v)$. Here, $\lambda$ and $\lambda^{-\nu}$ are the local expansion ratios parallel and perpendicular to the nematic director field and $R$ is a planar rotation matrix. Aharoni *et al.* further pose this problem as finding a coordinate transformation with constraints on the transformed metric tensor's eigenvalues (or equivalently, on the singular values of the transformation Jacobian). It is not known whether such coordinates always exist, and therefore, in the general case, this search becomes an optimization problem; finding coordinates whose Jacobian's singular values best match the desired ones at every point, *i.e.* minimize a "parameterization energy" functional:

$$E[\boldsymbol{u}] = \iint dist^2(J(\boldsymbol{u}), \mathcal{D})\, dS, \qquad (S2)$$

where $\boldsymbol{u}$ is the parameterization, $J(\boldsymbol{u})$ the Jacobian of the transformation from $\boldsymbol{u}$ to local Riemann normal coordinates, and $\mathcal{D} \subseteq GL(2, \mathbb{R})$ is the set of desired local Jacobians, namely, in our case, $\mathcal{D} = \left\{ U \begin{pmatrix} \lambda & 0 \\ 0 & \lambda^{-\nu} \end{pmatrix} V^T \mid U, V \in SO(2) \right\}$.

From this point of view, the inverse design problem is similar to other problems of finding optimal parameterizations of a surface, *e.g.* optimizing area preservation or angle preservation, many of which are highly applicative in computer graphics. For that goal, a discretized version of eq. (S2) is often used:

$$E[\boldsymbol{u}, \{L_t \in \mathcal{D}\}] = \sum_{t \in \Delta} A_t \|J_t(\boldsymbol{u}) - L_t\|^2, \qquad (S3)$$

where $t$ is taken over all mesh triangles, $A_t$ and $J_t$ are the areas and Jacobians respectively, and $L_t$ are the locally best-matching "allowed" Jacobians (optimization is performed over both $\boldsymbol{u}$ and $\{L_t\}$).

An efficient numerical scheme for optimizing eq. (S3) was presented in the seminal work of Liu *et al.* [S2]. Their approach, coined *Local/Global*, breaks the nonlinear optimization process into a sequence of iterations, each composed of two linear steps; a local step, in which the local Jacobian $J_t(\boldsymbol{u}^0)$ at every mesh triangle separately is approximated by the best-matching $L_t \in \mathcal{D}$; a global step, in which $\{L_t\}$ are assumed fixed, and a global parameterization $\boldsymbol{u}$ which minimizes $E$ is chosen (by fixing the $L_t$'s this becomes a linear problem). If $E[\boldsymbol{u}, \{L_t\}] < E[\boldsymbol{u}^0, \{L_t^0\}]$, this

process can be repeated until converging. If the convergence energy is zero (or a numerical approximation of it), than we have found a parameterization that everywhere follows the local singular value rule (or a numerical approximation of it). As a useful side effect, the singular value decomposition of the obtained $L_t$'s gives us the director field at every mesh triangle.

We employ the algorithm of [S2] using Matlab. We execute the local step by running singular value decomposition on the local 2×2 Jacobian at every mesh triangle, hence we get $J_t = USV^T$ where $U, V \in SO(2)$ and $S$ a diagonal matrix with eigenvalues ordered from large to small. We then define $L_t = U \begin{pmatrix} \lambda & 0 \\ 0 & \lambda^{-\nu} \end{pmatrix} V^T$. It is important to note that the solution at any triangle is completely independent of solution in neighboring triangles. Therefore, if we were to run the local/global algorithm on an arbitrary initial parameterization, we would typically get a very "noisy" limit parameterization, with director field (principal direction) changing on the scale of a single triangle. To avoid that, we run the entire local/global algorithm twice. The first run is made with $\lambda = 1$. This poses a very tight restriction on $\mathcal{D}$ as a result of the eigenvalue degeneracy, and the process typically converges to a parameterization with high residual energy (S3), commonly referred to as an "as rigid as possible" (ARAP) parameterization. We then stretch the entire coordinate space $\boldsymbol{u} \to \begin{pmatrix} \lambda & 0 \\ 0 & \lambda^{-\nu} \end{pmatrix} \boldsymbol{u}$ to align the local Jacobians, and run the local/global algorithm again with the desired $\lambda$ and $\nu$.

Numerical Inverse Design of 2D Curvature Tensor

It was shown in [S1] that a planar director field that is not homogeneous across the sheet's thickness, but rather changes from $\theta(u,v) + \Delta\theta(u,v)/2$ at one side to $\theta(u,v) - \Delta\theta(u,v)/2$ at the other side, where $|\Delta\theta| \ll 1$, induces on a nematic elastomer sheet the reference metric given by eq. (S1) (with $O(\Delta\theta)^2$ corrections) along with a reference curvature tensor

$$\bar{b} = \frac{\lambda^2 - \lambda^{-2\nu}}{2h} R[\theta] \begin{pmatrix} 0 & 1 \\ 1 & 0 \end{pmatrix} R[\theta]^T \Delta\theta + O(\Delta\theta)^3. \tag{S4}$$

In the thin limit, the ground state of a free elastic shell tends toward an isometry of its reference 2D metric tensor, which, among all isometries, minimizes the discrepancy between its curvature tensor and the shell's reference curvature tensor [S3]. For this reason, we first use the protocol described in the previous section to find a director field $\theta(u,v)$ that gives approximately the metric of our desired surface, and then use it as a given parameter in eq. (S4) to induce $\bar{b}$ that is as close as possible to our desired surface's curvature tensor $\tilde{b}$. Although we in fact search for $\Delta\theta(u,v)$ that minimizes the bending energy functional, we use the less accurate yet simpler process of finding a minimizer to the Frobenius norm $\iint \mathrm{Tr}\left[(\bar{b} - \tilde{b})^2\right] du\, dv$. Together with (S4), the minimizer is given by

$$\Delta\theta = \frac{2h}{\lambda^2 - \lambda^{-2\nu}} \mathrm{Tr}\left(R[\theta] \begin{pmatrix} 0 & 1 \\ 1 & 0 \end{pmatrix} R[\theta]^T \tilde{b}\right) = \frac{4h}{\lambda^2 - \lambda^{-2\nu}} \left(R[\theta]^T \tilde{b} R[\theta]\right)_{uv} \tag{S5}$$

where $(\cdot)_{uv}$ denotes the off-diagonal matrix element in the $\{u,v\}$ coordinates and $h$ is the thickness of the sheet.

We evaluate the curvature tensor $\tilde{b}$ of our desired triangulated surface at the centers of mesh triangles using standard methods [S4]. Combined with the director field $\theta$ at mesh triangles we get from the protocol at described in the previous section and the explicit recipe ($S5$), we obtain two planar director fields evaluated at mesh triangles, $\theta \pm \Delta\theta/2$, which we wish to impose upon the top and bottom surfaces of our thin nematic elastomer sheet. We thus create two masks – one for treating the top surface and one for treating the bottom surface (as described in Supplementary Text).

Materials

1,3-propanedithiol, 1,8-Diazabicyclo(5.4.0)undec-7-ene (DBU), butylated hydroxytoluene (BHT) and photoinitiator 2,2-dimethoxy-2-phenylacetophenone (DMPA) were purchased from Sigma Aldrich and used as received. Hydrochloric acid (HCl) was purchased from Fisher Scientific, and (tridecafluoro-1,1,2,2-tetrahydrooctyl)trichlorosilane (Silane-8174) was purchased from Gelest. Positive-tone photoresist, S1813 and epoxy resin D.E.R. 354 were purchased from Dow Chemical Company. Poly(dimethylsiloxane) (PDMS) precursor and curing agent, Sylgard ® 184, were purchased from Dow Corning. LC monomer, 1,4-bis-[4-(6-acryloyloxy-hexyloxy)benzoyloxy]-2-methylbenzene (RM82), was purchased from Wilshire Technologies Inc. and used without further purification.

Synthesis of LC Precursors

$10g$ of RM82 and $3.14g$ of 1,3-Propanedithiol (PDT) (mol ratio of RM82 : PDT=1:2) were dissolved into $100mL$ $CH_2Cl_2$ with rigorous stirring. 2 drops of 1,8 Diazabicycloundec-7-ene (DBU) were then added as catalyst. The reaction mixture was stirred at room temperature overnight.

After reaction, the mixture was washed with diluted HCl twice (1M first and 0.1M 2nd), followed by washing with DI water once. The $CH_2Cl_2$ solution was then dried with $MgSO_4$ for 30min, and filtered afterwards. The final product (RM82-PDT) was collected as viscous liquid after evaporating the solvent under vacuum.

Fabrication of Epoxy 1D Channels

Masters with 1D channels were fabricated from photoresist S1813 using direct laser writing (Heidelberg DWL 66+). Glass substrates were pre-cleaned by rinsing with acetone three times, followed by drying with an air gun. A thin layer of S1813 (thickness $\sim 2\mu m$) was spin-coated (4000rpm for 40s) on the clean glass substrate, followed by prebaking at $110°C$ for 1 min. The S1813 photoresist layer was then exposed by direct writing laser (365nm UV light) with a designed pattern. After laser writing, the sample was developed by photoresist developer (MF319) to obtain the final S1813 pattern.

A PDMS mold was replicated from the S1813 master following the procedure reported in [S5]. To create 1D channels with planar anchoring to align LCMs, the PDMS mold was replicated to epoxy (D.E.R. 354). One drop ($\sim 10\mu L$) of D.E.R. 354 liquid was first placed on a clean glass, followed by covering the uncured epoxy liquid with the patterned PDMS mold. The epoxy liquid

was then filled into the PDMS pattern through capillary infiltration. The D.E.R. 354 along with the PDMS mold was exposed to 365nm UV light with a dosage of 15,000 mJ/cm$^2$ to fully cure the epoxy. The final replica of epoxy was obtained by carefully peeling off the PDMS mold from the cured epoxy pattern.

Surface Coating of Epoxy Patterns

The as-prepared epoxy patterns were cleaned with a UVO cleaner (Jelight Company, Model 144AX) for 15min to slightly oxidized the epoxy surface. The samples were then moved to a vacuum chamber for fluoro-silane treatment through a chemical vapor deposition (CVD) process. A small drop ($\sim 5\mu L$) of fluoro-silane (Silane-8174) was placed at the center of the vacuum chamber surrounding by epoxy samples with similar distance towards the center droplet. As the chemical vapor pressure of the silane is dependent on its diffusion length, this would ensure a relatively uniform coating density on the epoxy samples. The chamber was then vacuum pumped for 25 min, and the final fluoro-silane coated (F-coated) samples were kept in a dry place for future use. The primary objectives of fluoro-silane coating are two folds: 1) F-coated epoxy gives planar anchoring towards the mixture of RM82 and RM82-PDT. 2) F-coating is well-known as low surface energy material for easy boundary delamination to separate LCE films from the epoxy patterns.

Preparation of LCE Sheets

The F-coated epoxy substrates were used as both the top and the bottom surfaces to construct the LC cells; they were aligned under an optical microscope with a mismatch $< 10\mu m$. The top and bottom channels were individually designed to construct a small twisting angle across film thickness following theoretical calculation. The thickness of the LC cells was controlled using Mylar (either 150 or $100\mu m$ thick) as spacer. After aligning the two epoxy patterns, the two substrates were then fixed by epoxy glue on the two edges of the sample.

To prepare LCM precursors, RM82 and RM82-PDT were mixed with a molar ratio 1:1 in a 5 mL vial, and ~1 wt% of DMPA was added as photoinitiator. Due to the high active nature of thiol-acrylate reaction, 0.1 wt% free radical inhibitor (BHT) was added to avoid polymerization before UV exposure. The mixture was melted on a hot stage at $100°C$ with vigorous stirring until fully mixed. To prevent reaction between RM82 and RM82-PDT while heating, short mixing time is preferred (1-2 min).

After full mixing, the mixture of LCM precursor was then infiltrated into the epoxy LC cell on the hot stage at $100°C$ through capillary force. The LC cell was then moved to a $60°C$ hot stage and annealed for 45 min to achieve LC alignment. The alignment of LCM in the cell was confirmed by polarized optical microscope (POM). LC cell was then exposed to UV irradiation (365nm, Newport model 97436-1000-1, Hg source) at a dosage of 1000-10K mJ/cm$^2$, and the final LCE sheets were obtained by peeling the thin film off the epoxy patterns.

Characterization

*Differential Scanning Calorimetry (DSC):* DSC tests were performed on a TA Instruments

Q2000 with an aluminum hermetic crucible. Sample was heated and cooled with ramping rate $10°C/min$ for three cycles, and data from the 2nd cycle was reported. Oligomer mixture of RM82/RM82-PDT was scanned from $120°C$ to $-40°C$, and the crosslinked LCE film was scanned from $180°C$ to $-30°C$. For the DSC measurement of the oligomer mixture, 0.1 wt% of BHT was added to avoid thermal polymerization during heating and cooling cycles, and the final data showed the three cycled curves are mostly identical. For the measurement of crosslinked LCE film, a polydomain sample was used and cut into small pieces to be contained in the aluminum crucible without mechanical pressing, and the DSC data showed a weak nematic to isotropic phase transition near $130°C$. This phase transition was also seen under POM during the measurement of actuation strain under heating. It should be noted that after the phase transition, the nematic phase still holds a significant amount of order parameter that gradually decreases with the increasing temperature till $180°C$. DSC also found a glass transition temperature of the LCE film around $0°C$, indicative of elastomeric nature of the crosslinked membrane.

*Gel Permeation Chromatography (GPC):* GPC was performed on a Tosoh EcoSEC GPC system with tetrahydrofuran as eluent at a flow rate of 1 mL/min. To ensure the separation of the oligomers, RM82-PDT sample was permeated through a combination of three single-pore gel columns (Tosoh). And the molecular weight of polymers in RM82-PDT was calculated by using polystyrene standards (Sigma Aldrich), as shown in Fig. S2.

*Tensile Testing:* Tensile testing was performed on an Instron machine (Instron Model 5564, with a 10 N load cell) under displacement control (10 mm/min). Rectangular samples (20 mm × 10 mm × 0.15 mm) were cut from a uniformly aligned samples and tested with the LC director either parallel or perpendicular to the displacement direction. As shown in Fig. S4b, the modulus of LCE is much higher in the director direction comparing to that in the perpendicular direction.

*Polarized Optical Microscopy (POM):* LC alignment was characterized by a motorized optical microscope (Olympus BX61) equipped with crossed polarizers using CellSens software. Heating and cooling of samples were performed in ambient air on a Mettler FP82 and FP90 thermo-system hot stage. POM was also used to monitor dimensional change of monodomain LCE films for the measurement of actuation strain. The actuation strain was calculated from the dimensional change of a small rectangular piece of monodomain LCE film floating on a thin layer of silicon oil upon heating on the hot stage. Images were taken with $10°C$ temperature intervals at a heating rate of $1°C/min$. Dimension of the films were directly measured from the images using CellSens software.

*LCE Actuation Characterization:* For thermal actuation, samples were heated to $180°C$ at a heating rate $\sim 50°C/min$ in ambient air on a hot plate (IKA C-Mag HS7) and cooled to room temperature naturally. Films were typically heated and cooled for several cycles (3-4 times) to release residual stress before image and video capture. Instead of ambient light, directional light was used and carefully adjusted to minimize surface iridescent reflection that comes from the micro-channels on the sample surface. Images and videos of the actuating LCE sheets were taken by a digital camera.

*Scanning Electron Microscopy (SEM) and Atomic Force Microscopy (AFM) Imaging:* SEM imaging was performed on a dual beam FEI Strata DB 235 Focused Ion Beam (FIB) SEM instrument with 5KV electron beam. AFM imaging was performed on a Bruker Dimension Icon

AFM under tapping mode.

*Measurement of 3D topography:* The 3D topography of the LCE face membrane (Fig. S7) was quantitatively measured by a Keyence VHX-5000 microscope. The 3D scan was taken while the LCE sample is heated to $180°C$ using a hot stage.

**Supplementary text**

<u>LC anchoring control</u>

LC anchoring on 1D channels is highly affected by the channel width and depth. The anchoring energy density follows Berreman's model [S6]:

$$\rho_{max} = \frac{\pi^2 K A^2}{\lambda^3} \qquad (S6)$$

where $K$ is the average elastic constant of LCs, $A$ and $\lambda$ are the depth and pitch (width + spacing) of the channel, respectively. $\rho_{max}$ is the anchoring strength of LCs on the 1D channels representing the potential of reorientation of LCs from other alignment directions to the channel direction. Our previous method [S7] created micro-channels through photolithography with $A = 1.5\mu m$ and $\lambda = 4\mu m$, and they worked nicely to create various topological defects. However, in our new LC oligomer system, it is found to be less effective: numerous unwanted defects are generated during the annealing process that strongly impedes uniform alignment. We suspect this to be attributed to the non-sufficient surface anchoring energy, and the sharp edges and corners of the channel pattern that could trap spurious topological defects. To circumvent this, we specially design our new channel patterns with smaller channel width and smoother surface topography, enabled by the advanced direct laser writing technique. Firstly, we spin-coated a $\sim 2\mu m$ thick positive-tone photoresist (S1813) onto a flat substrate. The film thickness was designed to be thicker than the desired channel depth such that only top of the photoresist will be developed to avoid the sharp bottom corners in the channel pattern. Secondly, we carefully adjusted the exposing laser power to secure the UV light only penetrated a small depth into the photoresist film. Because the uneven UV light distribution in the thickness of the photoresist layer that the top layer absorbs light more than the bottom layer, the obtained channel pattern would have a smooth surface topography, which was confirmed by SEM (Fig. S6a) and AFM (Fig. S6b) measurements. As shown from the SEM and AFM images, neither the top or the bottom of the channels has sharp edges and $\lambda$ is reduced to $2\mu m$, which from eq. $(S6)$ would increase surface anchoring energy by 8 times. Therefore, with the reduced λ, LCs can achieve strong surface anchoring on much smaller surface corrugation/channel depth. Our finding shows that 600nm is the near-optimized channel depth in consideration of LC surface anchoring and lithographic fabrication.

<u>Inducing Reference Curvature</u>

Even though inverse calculation from theory gives 2D director patterns, the parallel alignment of LC molecules across the LCE film results in multiple isometries of the shape

transformation. As mentioned in the main text, we augment our design with imposing a slight twist of LC molecules from the top to the bottom of the film. The calculation of $\Delta\theta$, the angle difference between the top and bottom director fields for which the target shape minimizes bending energy among its isometries, is detailed in the Materials and Methods section. For the face sample in main-text Fig. 4, with film thickness $100\mu m$, we get an angle difference within the range $-3° < \Delta\theta < 3°$ over the entire sample.

In experiment, however, inducing reference curvature needs to serve several purposes (as discussed in the main text), one of which is to help the initially flat sheet buckle in the correct direction and dynamically move toward the preferred equilibrium shape. For that reason, we made several samples, in which we overexpressed the target curvature at every point ($\Delta\theta \to 2\Delta\theta$ over the entire sample) or just around the edges ($\Delta\theta \to \left(1 + 5e^{-d/2mm}\right)\Delta\theta$, with $d$ the distance to the boundary), to check whether we can get better dynamical behavior and still have the equilibrium not far from our target shape. We noticed that the overly curved LCE films often deformed into relatively irregular shapes, far away from theoretical prediction. Films with the calculated $\Delta\theta$, for most of times, were able to find a route to the correct isometry, demonstrating the face shape. We show our best results in Fig. 4 and Supplementary Movie S4. We therefore conclude that using the direct calculation of $\Delta\theta$ described in Materials and Methods is preferable.

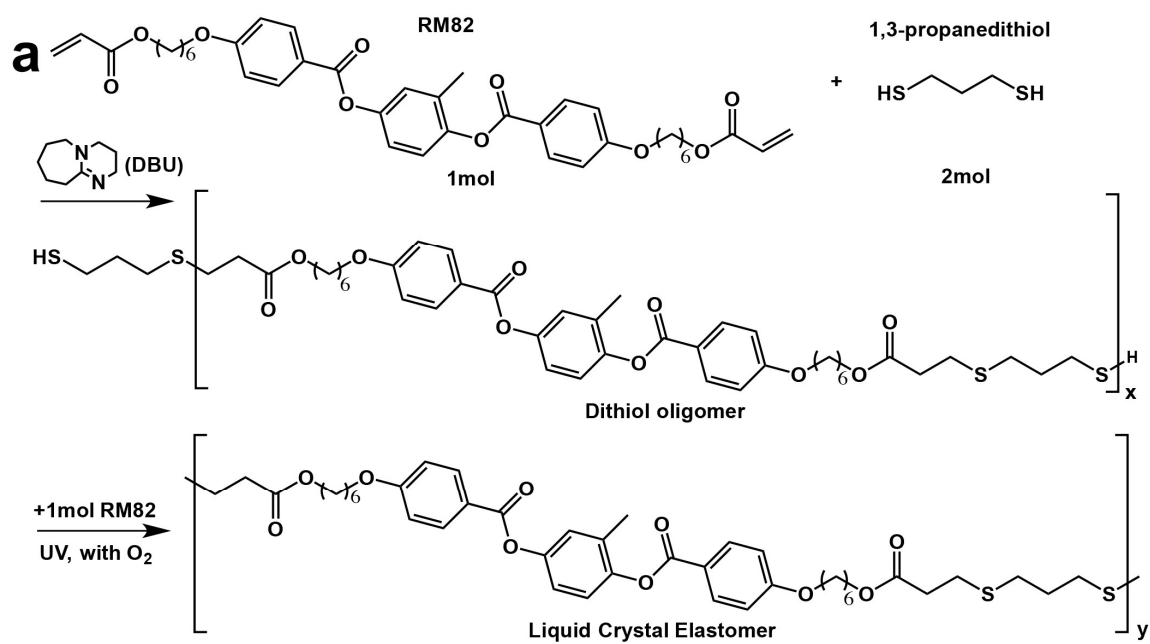

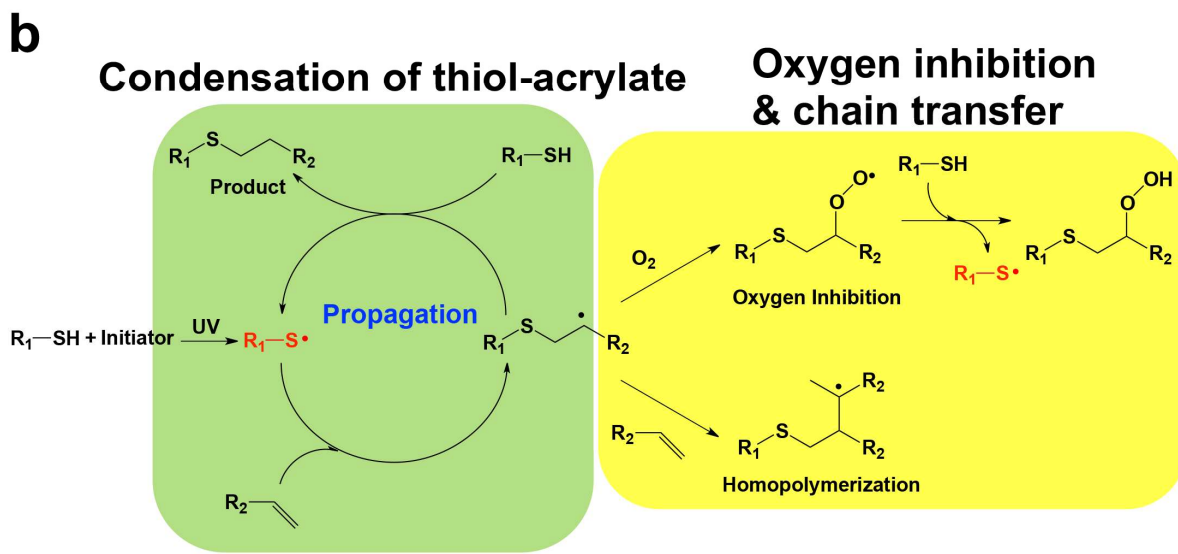

**Fig. S1** Schematic illustration of (a) chemical reactions used in this work. (b) oxygen mediated thiol-acrylate "click" reaction. The oxygen here serves as a chain transfer agent that transfers the active free radical from the homopolymerization of acrylates to the condensation reaction of thiol-acrylate, which effectively lowers the crosslinking density of the system and results in an elastomeric polymer network.

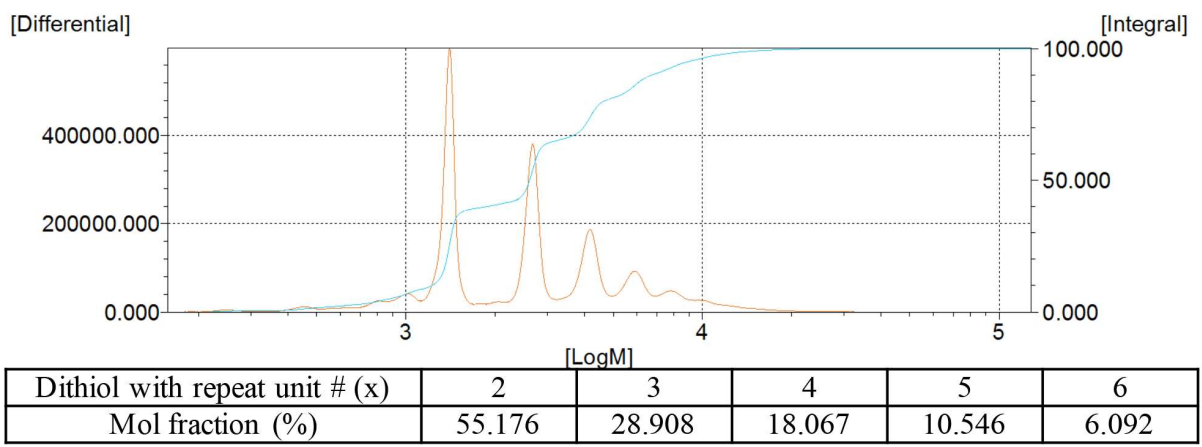

| Dithiol with repeat unit # (x) | 2 | 3 | 4 | 5 | 6 |
|---|---|---|---|---|---|
| Mol fraction (%) | 55.176 | 28.908 | 18.067 | 10.546 | 6.092 |

**Fig. S2** Gel permeation chromatography (GPC) measurement of RM82-PDT. The fraction of each oligomer with different chain length is estimated from the area of the corresponding peak. One can clearly see the major part of the oligomer is short chain oligomers with repeat unit number x =2, 3, 4.

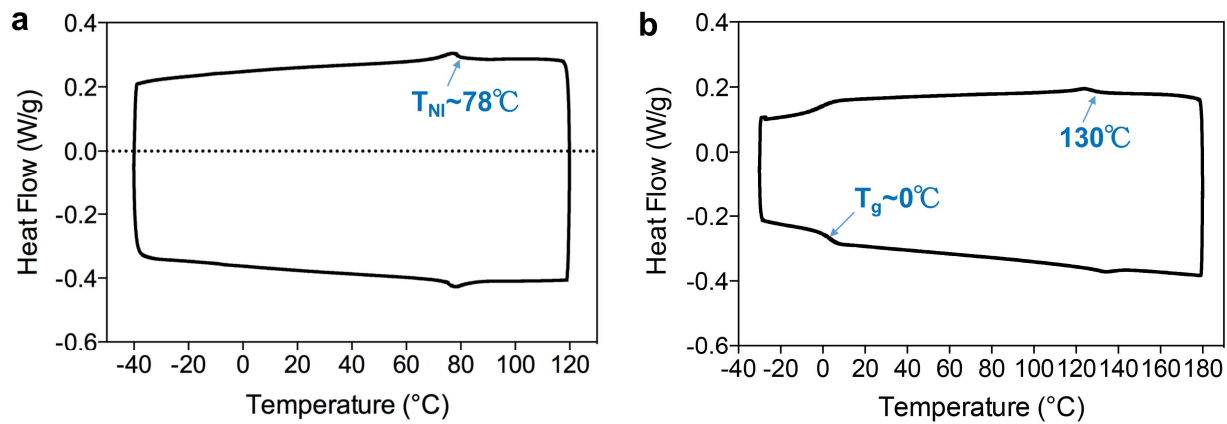

**Fig. S3** Differential scanning calorimetry (DSC) measurements of LCM mixture (a) and the crosslinked LCE (b). (a) The nematic-isotropic phase transition temperature ($T_{NI}$) is found ~ 78ºC, and the LCM mixture shows no crystallization until -40 ºC. (b) The film shows an elastomeric characteristic with a $T_g$ ~ 0 ºC, and an nematic-isotropic phase transition ~ 130 ºC.

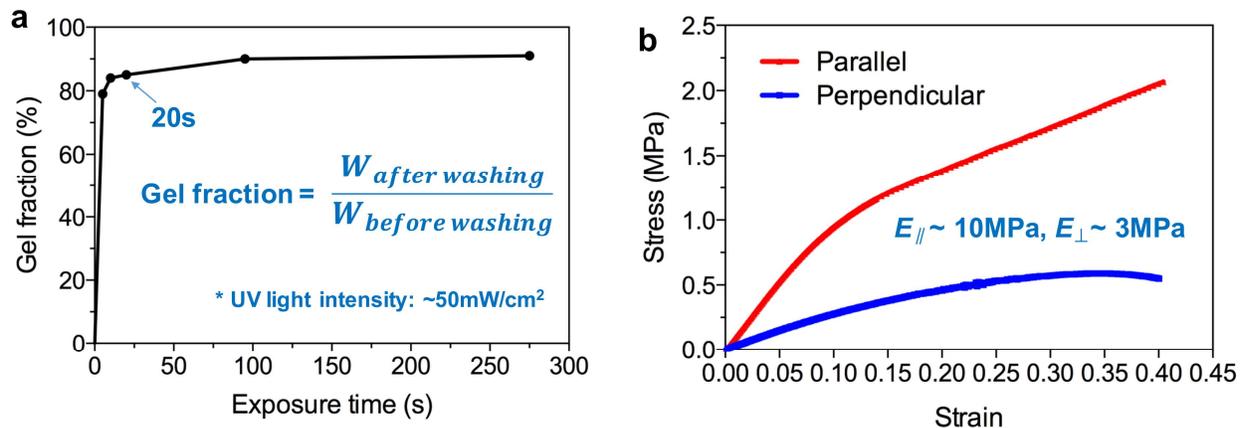

**Fig. S4** Characterization of the thiol-acrylate LCE. (a) Polymerization speed of the thiol-acrylate "click" reaction measured from gel fraction of the LCE film exposed to UV at different exposure time. (b) Mechanical measurement of stress-strain property of LCE film.

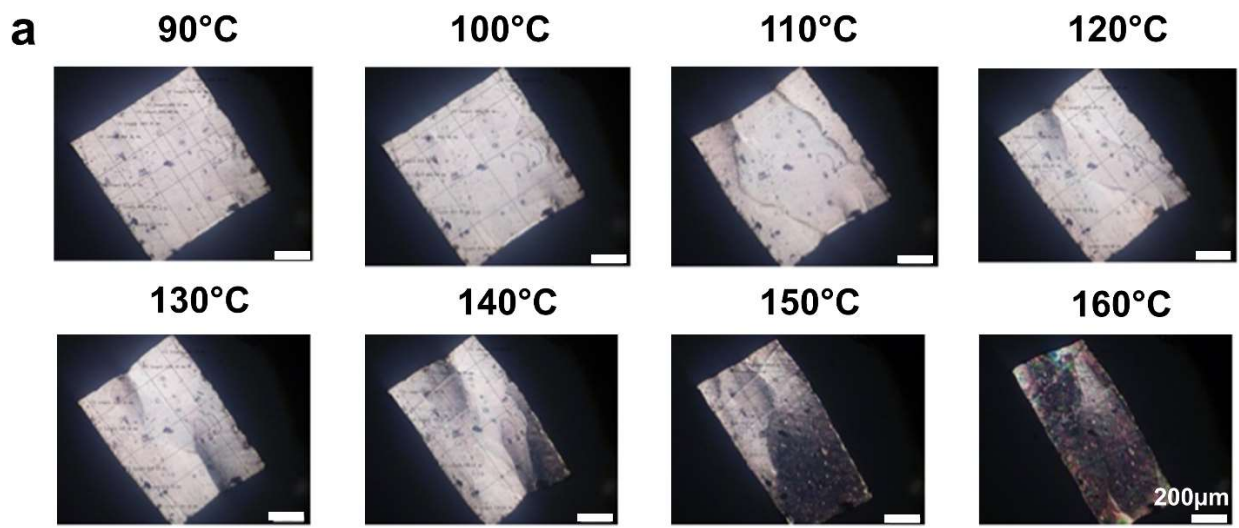

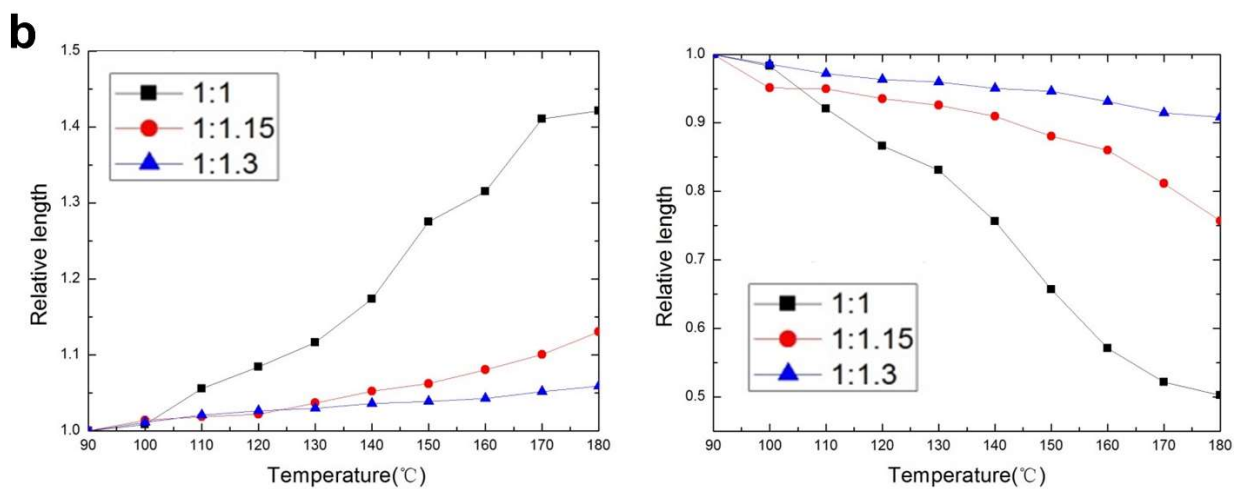

**Fig. S5** Characterization of LCE film for thermal actuation. (a) POM images show the deformation of a uniformly aligned LCE film under heat. (b) Thermal actuation of LCEs with different crosslinking. Black, red and blue dots represent different molar ratio of RM82:dithiol from 1:1, 1:1.15 to 1:1.3, respectively.

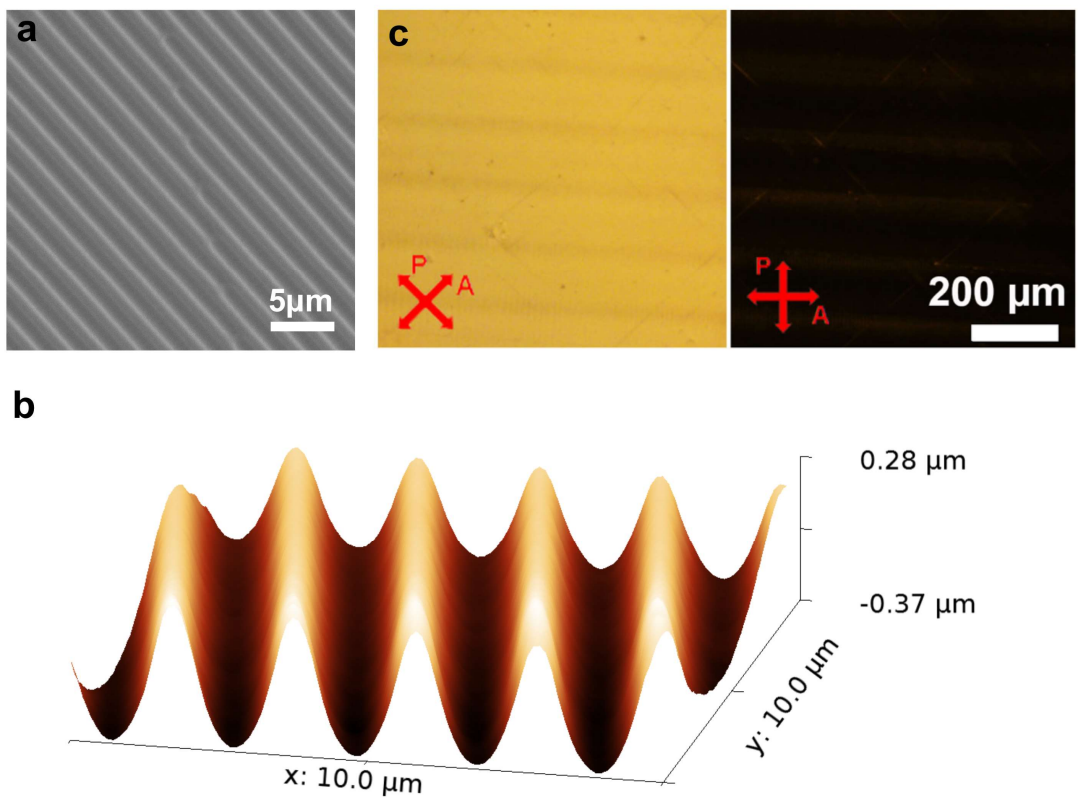

**Fig. S6** Characterization of 1D channels and their LCM alignment property. (a) SEM image of 1D micro-channels. (b) AFM measurement of 1D micro-channles. (c) LCMs on top of 1D channels at 45º(left) and 0º(right) polarizer angle. Uniform LC alignment is strongly imposed by the 1D channels without any defect.

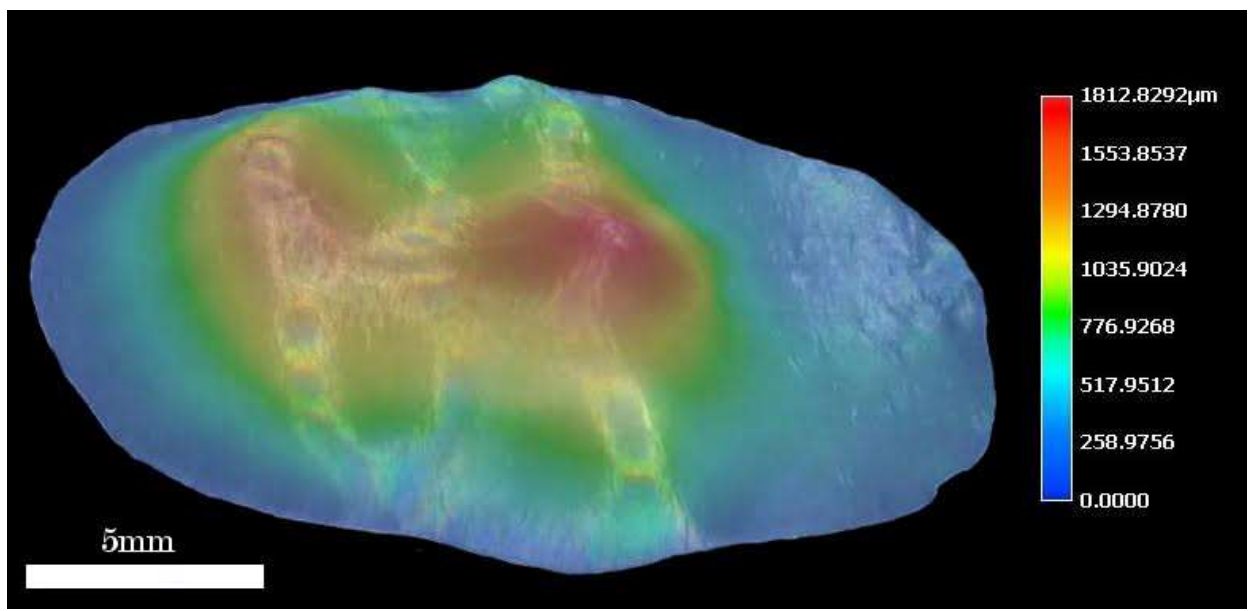

**Fig. S7** Topography of the LCE face membrane at $180°C$. The 3D shape is presented using the local height of the buckled film. Colors corresponding to values of the height function are indicated by the color bar at the right.

**Movie S1**

LCE-type deformation visualized with Tissot's indicatrices. When the cylinder-shaped sheet is actuated, small circles transform into identical ellipses with principal axes $\{\lambda, \lambda^{-\nu}\}$, locally aligned with the director field explicitly calculated by Aharoni *et al.* in [S1]. The deformation globally transforms the cylinder into a sphere.

**Movie S2**

LCE sheet programmed to obtain constant positive Gaussian curvature, heated to $180°C$. Heating rate is $\sim 100°C/min$, video speed is $3.5\times$ real time.

**Movie S3**

LCE sheet programmed to obtain constant negative Gaussian curvature, heated to $180°C$. Heating rate is $\sim 100°C/min$, video speed is $3.5\times$ real time.

**Movie S4**

LCE sheet programmed to obtain the shape of a leaf, heated to $180°C$. Heating rate is $\sim 100°C/min$, video speed is $3.5\times$ real time.

**Movie S5**

LCE sheet programmed to obtain the shape of a face, heated to $180°C$. Heating rate is $\sim 100°C/min$, video speed is $3.5\times$ real time.



\end{document}